\documentstyle[prb,aps,epsfig,multicol]{revtex}

\begin{document}
\title{Electronic structure of the pyrochlore metals
Cd$_{2}$Os$_{2}$O$_7$ and Cd$_{2}$Re$_{2}$O$_7$}
\author{D.J. Singh}
\address{Code 6391, Naval Research Laboratory, Washington, DC 20375}
\author{P. Blaha and K. Schwarz}
\address{Institut f\"ur Physik und
Theoretische Cheme, TU Wien, A-1060 Wien, Austria}
\author{J.O. Sofo}
\address{Centro Atomica Bariloche and Instituto Balseiro,
Comision Nacional de Energia Atomica, 8400, Bariloche, Argentina}
\date{\today}
\maketitle

\begin{abstract}
First principles density functional calculations within the local spin
density approximation (LSDA) and the generalized gradient approximation (GGA)
are reported for pyrochlore Cd$_2$Os$_2$O$_7$ and Cd$_{2}$Re$_{2}$O$_7$.
The transition metal $t_{2g}$ manifolds
are found to be well separated in energy from the O $2p$ bands and from
the higher lying $e_g$ and Cd derived bands.
The active electronic structure in the $t_{2g}$ manifold near the 
Fermi energy, $E_F$ is found to be significantly modified by spin orbit
interactions, which we include. Both materials show semi-metallic
band structures, in which the $E_F$ lies in an pseudogap.
The band structure of Cd$_2$Os$_2$O$_7$ near $E_F$ is dominated by
very heavy hole and electron bands, though at $E_F$ the electron sections
are lighter.
Cd$_{2}$Re$_{2}$O$_7$
has heavy hole bands but moderate mass electron states. The results
are discussed in terms of measured transport and thermodynamic properties
of these compounds as well as the very different ground states of
these two materials.
\end{abstract}
\begin{multicols}{2}

\section{Introduction}

Investigations of transition metal oxides with metal-insulator
transitions in their phase diagrams have revealed a remarkable range
of interesting, complex and often unanticipated phenomena, including
high temperature superconductivity, various charge, spin
and orbital ordered states, triplet superconductivity, giant magnetoelastic
effects, and heavy Fermion behavior. A good example is provided by
the perovskite manganites. Although known for several decades, these
compounds were re-investigated in detail during the last few years because
of interest in the colossal magnetoresistive effect, which is basically
a metal-insulator transition occurring at or near the magnetic ordering
temperature for some compositions. What has been revealed is a complex
phase diagram including charge, spin and orbital ordered phases resulting
from an interplay of strong correlations, strong lattice coupling,
and band structure effects.

The pyrochlore oxide, Cd$_2$Os$_2$O$_7$ is metallic at room temperature,
but undergoes a metal-insulator transition at 226 K. \cite{sleight,mandrus}
Upon cooling through the transition, the resistivity, $\rho(T)$,
crosses over from
a metallic temperature dependence to an insulating ($\rho$ strongly increasing
as $T$ is reduced) shape, though it does not fit a standard activated form.
Instead, Sleight and co-workers \cite{sleight} fit $\rho(T)$ to a form
consistent with a second order transition where the gap goes smoothly
to zero at the critical temperature. Consistent with this, Mandrus and
co-workers \cite{mandrus} report that the specific heat anomaly resembles
a mean field transition and shows neither latent heat nor hysteresis.
They also reported extensive crystallographic data, showing very little
coupling of the transition to lattice degrees of freedom
and confirming the purely electronic character of the transition.
The specific heat shows a very small but possibly still finite
Sommerfeld coefficient (note that the reported lattice part does not fit
a simple Debye model complicating the analysis);
$\gamma \sim 1 ~$ mJ/{mole K$^2$} in the range 2K to 4K and the
low $T$ susceptibility is high for an ordered antiferromagnetic
insulating phase. The thermopower above the transition is hole-like and
small.
Just below the transition, the thermopower increases to $+40 ~ \mu V / K$,
but then decreases and crosses zero.
A very large electron-like thermopower, up to $S = -300 ~ \mu V / K$ is
seen well below the transition, \cite{sleight,mandrus}
This is not expected from the
simplest picture of a
fully gapped Fermi surface due to a spin density wave. On the other hand,
supposing that the metal-insulator
transition is magnetic in character, the ground
state could be very complicated especially in view of the
strong geometric frustration of the pyrochlore lattice and the fact that the
interactions are most likely antiferromagnetic.
The related compound, Cd$_2$Re$_2$O$_7$, which differs from Cd$_2$Os$_2$O$_7$
by one electron per transition metal, was recently reported to be
superconducting at $\sim$1K with a substantial diamagnetic signature.
\cite{hanawa,jin}
It displays metallic properties up to 200K
where an apparently isostructural transition of unknown character occurs.
\cite{hanawa,donohue,blacklock,subramanian}
Like Cd$_2$Os$_2$O$_7$, this transition shows signatures in
electrical transport and susceptibility, though the details differ.
Interestingly, the low temperature linear specific heat coefficient
for Cd$_2$Re$_2$O$_7$, $\gamma$=30 mJ/molK$^2$ (Ref. \onlinecite{hanawa})
is similar to the value $\gamma\sim20$mJ/molK$^2$ estimated for the
high temperature phase of Cd$_2$Os$_2$O$_7$ (Ref. \onlinecite{mandrus})
in spite of the difference in electron count and the very different
low temperature properties.

Here we report the electronic
structures of these two compounds as
calculated within the local density
approximation (LDA) and the generalized gradient approximation (GGA),
using the general potential linearized augmented planewave (LAPW) method
including spin-orbit, which we find important.
\cite{singhlapw,lapwnote,blahalapw,weilapw,conv}
The calculations were done with well converged zone samplings and basis
sets (including local orbitals to relax linearization errors and treat
semi-core states). The experimental crystal structure at 180K of Mandrus
and co-workers, ({\it i.e.} $a$=10.1598\AA~and $x$(O1)=0.319)
was used throughout for Cd$_2$Os$_2$O$_7$. \cite{mandrus}
The calculated Hellman-Feynman force on the symmetry unconstrained
structural parameter was found to be small with this choice supporting
the experimental crystallography. Total energy minimization,
within the LDA yields a value $x$(O1)=0.3198.
Donohue and co-workers \cite{donohue} refined the crystal structure
of Cd$_2$Re$_2$O$_7$ obtaining $a$=10.219\AA~and $x$(O1)=0.309.
However, this yields rather short Re-O bond lengths, and we note that
the refinement is difficult because the compound contains heavy elements
and the presence of Cd mandated the use of X-rays rather than neutrons.
We performed a relaxation of the internal coordinate within the LDA keeping
the lattice parameter fixed at its no doubt reliable experimental value.
We obtained $x$(O1)=0.316 in this way. The energy of this relaxed structure
was 0.35 eV / cell (4 Re atoms) lower in energy than with the reported
value of $x$(O1)=0.309, which is well outside the normal LDA errors.
The Raman active $a_{1g}$ phonon frequency corresponding to this coordinate
is calculated as 463 cm$^{-1}$. The corresponding frequency for 
Cd$_2$Os$_2$O$_7$, also calculated in the LDA is almost the same,
$\omega$=459 cm$^{-1}$, indicating that the Re compound is not
softer than the Os regardless of the unusual Re valence, at least as
measured by this particular phonon mode.

In many aspects, the calculations we present are like those
presented by Mandrus and co-workers for Cd$_2$Os$_2$O$_7$.
However, there is one important
difference. We find that spin orbit interactions are significant
because of the presence of heavy elements in the structure, and so
include them by the usual second variational approach. The inclusion
of spin-orbit substantially changes the electronic structure near the
Fermi energy. In addition, we allow for
magnetism in Cd$_2$Os$_2$O$_7$, which we investigated
using self-consistent unconstrained and fixed spin moment calculations.
These were done by the method described in Ref. \onlinecite{ashk}
Not surprisingly, our calculated non-spin-polarized
band structure for Cd$_2$Os$_2$O$_7$ {\it without including
spin orbit} is practically identical to that given by Mandrus and co-workers
and so is not displayed here. Significantly,
it is very metallic and has four Os $d$ derived
bands crossing the Fermi energy $E_F$ and other bands. It is
difficult to envisage what kind of instability might make it insulating.
In the remainder of this report we discuss the band structure {\it
including spin orbit}, which while similar in many basic aspects differs
significantly near $E_F$.

\section{Band Structure}

The crystal structure of Cd$_2$Os$_2$O$_7$ features Os ions (nominally
Os$^{+5}$, 5$d^3$) at the center of O octahedra.
Cd$_2$Re$_2$O$_7$ is similar but with one less 5$d$ electron per transition
metal atom.
Within an ionic model, assuming nominal charges, one expects a 
manifold of occupied O $2p$ bands, followed by a partially filled
transition metal
$t_{2g}$ manifold and a higher lying unnoccupied set of $e_g$ bands.
Furthermore, since the actual transition metal site symmetry is
weakly rhombohedral (due to the second neighbor coordination), a further
crystal field splitting of the $t_{2g}$ manifold is possible.

\begin{figure}[tbp]
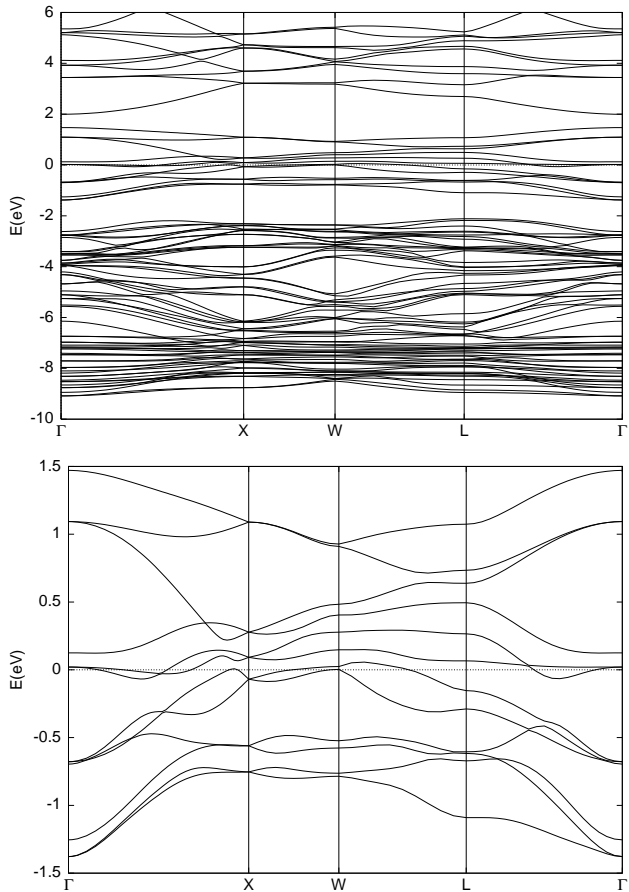

\centerline{\epsfig{file=band-so.epsi,angle=270,width=0.95\linewidth}}
\vspace{0.10in}
\centerline{\epsfig{file=band-so-small.epsi,angle=270,width=0.95\linewidth}}
\vspace{0.15in}
\setlength{\columnwidth}{3.2in} \nopagebreak
\caption{
LDA paramagnetic band structure
of Cd$_2$Os$_2$O$_7$. Spin-orbit interactions were included via
a second variational step. The horizontal reference denotes
$E_F$. The bottom panel is a blow-up around
$E_F$ showing the Os $t_{2g}$ manifold.}
\label{ldabands}
\end{figure}

\begin{figure}[tbp]
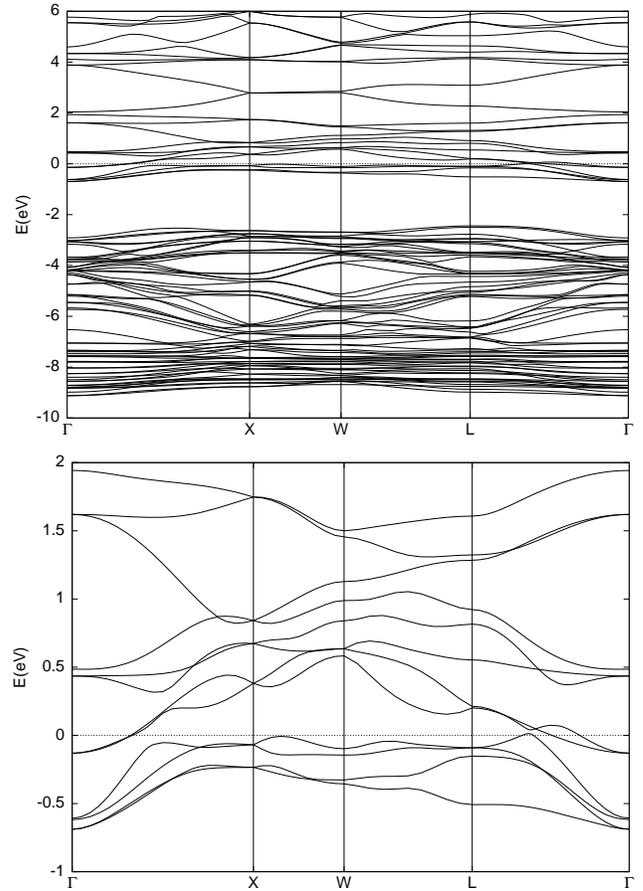

\centerline{\epsfig{file=band-re.epsi,angle=270,width=0.95\linewidth}}
\vspace{0.10in}
\centerline{\epsfig{file=band-re-fine.epsi,angle=270,width=0.95\linewidth}}
\vspace{0.15in}
\setlength{\columnwidth}{3.2in} \nopagebreak
\caption{
LDA paramagnetic band structure
of Cd$_2$Re$_2$O$_7$. Spin-orbit interactions were included via
a second variational step. The horizontal reference denotes
$E_F$. The bottom panel is a blow-up around
$E_F$ showing the Re $t_{2g}$ manifold. The heavy bands cross $E_F$
yielding hole sections of Fermi surface, but these are not along the
symmetry lines shown.
Note the non-rigid band behavior
relative to Fig. \ref{ldabands}.}
\label{re-bands}
\end{figure}

\begin{figure}[tbp]
\centerline{\epsfig{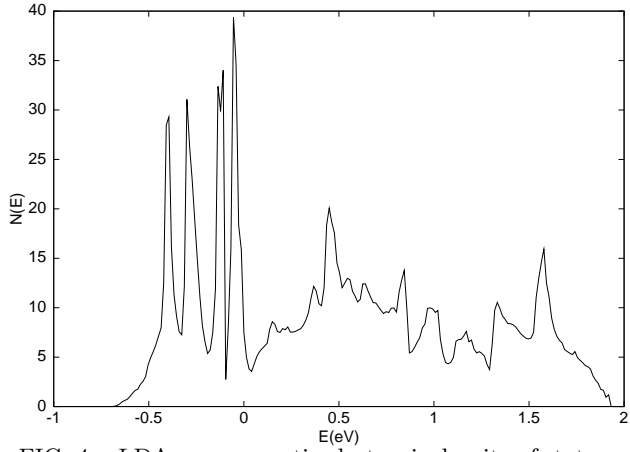}}
\setlength{\columnwidth}{3.2in} \nopagebreak
\caption{
LDA paramagnetic electronic density of states of Cd$_2$Re$_2$O$_7$
for the $t_{2g}$ manifold on a per unit cell (4 Re atoms)
basis. Spin-orbit is included.}
\label{re-dos}
\end{figure}

The calculated LDA band structure of Cd$_2$Os$_2$O$_7$
is shown in Fig. \ref{ldabands}, while that of Cd$_2$Re$_2$O$_7$
is in Fig. \ref{re-bands}.
The corresponding non-spin-polarized
densities of states (DOS) near $E_F$ are shown in
Fig. \ref{ldados} and \ref{re-dos}.
The GGA band structures (not shown) are very similar.
As shown by the band structure the $t_{2g}$ and
$e_g$ manifolds are cleanly
separated from each other, and from the O $2p$ bands by clean gaps, as in
the ionic model.
However, despite this there is quite strong hybridization between Os $d$ and
O $p$ states, reflecting the covalent
tendency of $5d$ transition metals relative to
$3d$ oxides.
Although the actual Os site symmetry is rhombohedral,
this part of the crystal field is weak and the band structure shows
no apparent further splitting of the $t_{2g}$ derived manifold.
It is exactly
half-filled and contains 12 bands (note that there are 4 Os ions per
unit cell). The $t_{2g}$ band width is 2.85 eV. Considering that the
effective Hubbard $U$ is likely 2 eV or less, based on the trends
for transition metal oxides, and noting the multichannel character of the
$t_{2g}$ manifold, Cd$_2$Os$_2$O$_7$ should not be classified as a 
strongly correlated material in the sense of having on-site Coulomb
correlations play a dominant role in the formation of the electronic
structure.
This is consistent with the observation of a high temperature
metallic phase. \cite{hub-note}

As mentioned, the $t_{2g}$ manifold, which consists of 12 bands, is
exactly half-filled. Unlike the scalar relativisitic band structure,
the band structure including spin-orbit shows a semi-metallic structure
in the sense that there is a gap between the sixth and seventh bands
throughout the Brillouin zone, but because of the dispersions it
is not an insulating gap. Instead, there are two electron-like
Fermi surfaces. One, from the lowest conduction band is a shell
around the $\Gamma$ point, while the other consists of ellipsoids
along the $\Gamma$-X lines. Corresponding to these electron surfaces
there are hole surfaces at the zone boundary around the W points.
This band structure can, at least conceptually, be made insulating
in two ways: (1) by increasing the gap between the sixth and seventh
bands or (2) by depressing the sixth band at the zone boundary and or
raising the, degenerate at $\Gamma$, seventh and eighth bands near the
zone center. The heavy masses of these narrow bands is consistent
with high thermopowers if a gap is opened. From the band structure, it
may be seen that the electron sheets of Fermi surface are lighter (higher
velocity) than the hole sections.
For the
paramagnetic metallic state,
the calculated average Fermi velocity is low reflecting these heavy 
bands, $<v_x^2>^{1/2}=<v_y^2>^{1/2}=<v_z^2>^{1/2}=5.9\times10^6$ cm/s

The calculated electronic DOS, $N(E)$ for Cd$_2$Os$_2$O$_7$
has a large peak just above $E_F$ with $N(E_F) = 9.24$ eV$^{-1}$
on a per formula unit (2 Os atom) basis. The GGA yields a somewhat
higher value $N(E_F) = 11.4$ eV$^{-1}$. However, while these numbers are
smaller than the scalar relativistic value of 12.7 eV$^{-1}$
reported by Mandrus and co-workers, \cite{mandrus}.
our LDA specific heat coefficient $\gamma$=22 mJ/molK$^2$
is still consistent within the range of 
electronic specific heats above the 
transition as estimated from experiment. \cite{mandrus}
This leaves little room for enhancement by electron phonon
interactions or beyond density functional many body correlations.

The band structure and corresponding DOS for Cd$_2$Re$_2$O$_7$ are
shown in Figs. \ref{re-bands} and \ref{re-dos}, respectively.
As may be seen from the upper panel of Fig. \ref{re-bands}, the large energy
scale electronic structure is like that of Cd$_2$Os$_2$O$_7$, and in 
particular the nominal ionic model is valid, with $E_F$ falling in an
isolated manifold of transition metal $t_{2g}$ states.
However, a closer
examination (lower panel of Fig. \ref{re-bands}) shows that the differences
from the Os compound are not at all well described by a rigid band model.
The band structure, like that of the Os compound is semi-metallic. In
particular, there is a clean pseudogap between the fourth and fifth
bands, and since there are eight $d$ electrons per cell the nominal
Fermi energy lies between them
The Fermi surfaces consist of nearly spherical, moderate mass
$\Gamma$ centered
electron pockets from the fifth and sixth bands in the $t_{2g}$ manifold
($m^*\sim1.2$),
and very heavy hole sections from near the zone boundary.
These enclose a total of 0.15 e/cell and an equal number of holes.
As may be seen,
this is less symmetric than the situation in Cd$_2$Os$_2$O$_7$.
However, because of the semi-metallic
character, which places $E_F$ near the band edges, the velocity is
still rather low, $<v_x^2>^{1/2}=7.5\times10^6$ cm/s.
Jin and co-workers \cite{jin} report that the Hall number is
quite $T$ dependent, but is electron-like at low temperatures. This
is consistent with our band structure, which has
both electron and hole sheets, but with much lighter electron sheets,
which will dominate the conductivity due to their higher velocities.

The calculated $N(E_F)$ for Cd$_2$Re$_2$O$_7$ is 5.3 eV$^{-1}$ per
formula unit, and is derived mainly from the heavy hole bands. This
corresponds to a bare band specific heat $\gamma$=12.4 mJ/molK$^2$.
Comparing with the measured value of 29.6 mJ/molK$^2$ (Ref. \onlinecite{jin})
one obtains an enhancement $(1+\lambda_{total})=2.4$ or
$\lambda_{total}=1.4$, a reasonable value for a known
superconductor. 

The plasma frequency for Cd$_2$Re$_2$O$_7$ from the calculated $N(E_F)$
and $<v_x^2>$ is $\hbar\omega_p$=1.3eV. Within Boltzmann tranport
theory for conventional metals, the slope of the intrinsic resistivity,
in the moderate temperature linear regime, is given by
$\partial\rho / \partial T=(8\pi^2\hbar k_B\lambda_{tr})/(\hbar\omega_p)^2$,
where $\lambda_{tr}$ is the transport electron phonon coupling, often
a reasonable approximation to the superconducting coupling $\lambda$.
Jin and co-workers \cite{jin} report approximately linear $\rho$ vs. $T$
from approximately 50-200K. 
However, the two reported samples differ by a factor of more than three
in $\partial\rho / \partial T$, providing a broad
range of $\lambda_{tr}$ from 0.75 to 2.5. This is problematic,
as according to the experimental data $\partial\rho / \partial T$
displays unusual though fine structure in this T range, and then changes
discontinuously to a value near zero at the transition. It is extremely
hard to imagine such a drastic change in $\lambda_{tr}$ originating from
the conventional mechanism, as it would suggest a high temperature phase
with almost no electron phonon scattering, which crosses over into
a moderate to strong coupled low temperature phase.

The bare susceptibility from the value $N(E_F)$ 
is $\chi_{B}=1.69\times10^{-4}$ emu/mol. The experimental
data show a large spread. Jin and co-workers \cite{jin} estimate
$\chi(0)$=5.4$\times10^{-4}$ emu/mol, while Hanawa and co-workers
\cite{hanawa} obtain 3.0 $\times10^{-4}$ emu/mol at low temperature,
rising to $\approx 5\times10^{-4}$ emu/mol above the 200K phase
transition. This corresponds to Wilson ratios $R_W$ from
0.74 to 1.3, which are low for a transition metal oxide,
but would indicate weak electron correlations in the 
presence of a moderate electron phonon coupling.

\section{Magnetism in C\lowercase{d}$_2$O\lowercase{s}$_2$O$_7$}

\begin{figure}[tbp]
\centerline{\epsfig{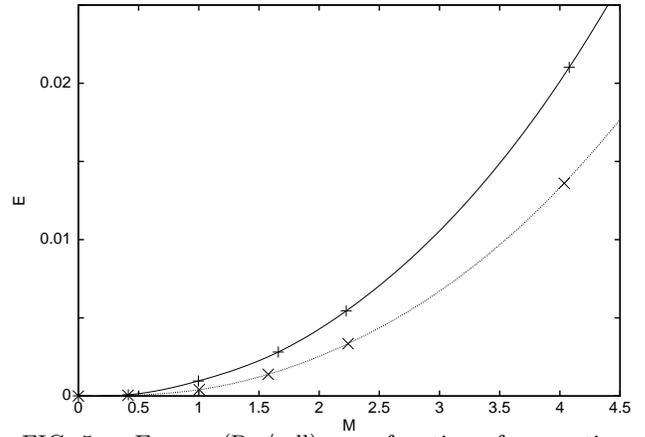}}
\setlength{\columnwidth}{3.2in} \nopagebreak
\caption{
Energy (Ry/cell) as a function of magnetization ($\mu_B$/cell) for
Cd$_2$Os$_2$O$_7$ from fixed spin moment calculations including spin orbit.
The solid (dotted) lines are for the LSDA (GGA).}
\label{fsm}
\end{figure}

It should be noted that the high value of $(N_F)$
obtained in Cd$_2$Os$_2$O$_7$ would lead to
a strong magnetic instability in a 3$d$ based material, but is marginal
here. This can be understood in terms of the expected lower Stoner
parameter $I$ 
in a 5$d$ compound.
The peak near $E_F$
derives from the flat practically dispersionless band that
lies just above $E_F$ over most of the zone. The relatively high $N(E_F)$
suggests the possibility of a Stoner instability against ferromagnetism.
We checked for this both within the LSDA and GGA using fixed spin moment
calculations, but found that Cd$_2$Os$_2$O$_7$ is predicted to be
stable against ferromagnetism, so there is not such an instability.
The calculated susceptibility is $\chi = 9.0\times10^{-7}$ emu/g in the LSDA.
This was determined from a fourth order fit of the energy {\it vs.}
moment for the small moment part of Fig. \ref{fsm}.
The bare Pauli susceptibility from $N(E_F)$ is
$\chi_{B} = 4.2\times10^{-7}$ emu/g, which
yields a Stoner enhancement of $(1-NI)^{-1}\sim2.2$. The GGA,
which sometimes overestimates the tendency towards magnetism in
4$d$ and 5$d$ oxides, places Cd$_2$Os$_2$O$_7$ closer to a magnetic
instability, with $\chi = 2.4\times10^{-6}$ emu/g and
$\chi/\chi_{B} = (1-NI)^{-1}\sim6$. Taking into account the
difference in the LDA and GGA values of $N(E_F)$, the GGA value of the
effective Stoner $I$ is 50\%
higher than the corresponding LDA value. This is reminiscent of the 
situation for Sr$_2$RuO$_4$, where LSDA calculations correctly
produce a paramagnetic state,
while the GGA
produces an incorrect ferromagnetic ground state, 
\cite{mazin,degroot}
due to an overestimated $I$.
In any case, Stoner ferromagnetism would rigidly exchange split the 
band structure, at least for small moments; as can be seen from the
band structure,
this would not result in an insulating electronic structure.

The band structure is quite isotropic (note {\it e.g.}
the similar dispersions around $E_F$
along $\Gamma$-X and $\Gamma$-L in Fig. \ref{ldabands}) and does
not display strong nesting. Thus there is no obvious preferred
wavevector to check in searching for a magnetic instability.
However, noting that there are four Os atoms in the unit cell, we
checked for a $\Gamma$ point antiferromagnetic instability, in which
two of the four Os atoms are spin up and two are spin down.
In principle, a state like this could produce an insulating band structure.
This is because the site dependent on-site exchange splittings
reduce hopping between opposite spin Os atoms, thus potentially
narrowing the bands enough to open the pseudogap between the sixth
and seventh bands producing a full band gap. While we find
that Cd$_2$Os$_2$O$_7$ is unstable against such an antiferromagnetic ordering
in the GGA without spin orbit, we find that spin-orbit favors
the paramagnetic state. In the LSDA an antiferromagnetic
ordering of this type does not occur. Within the GGA, including spin-orbit,
we find that
the material is on the borderline of an
instability against this antiferromagnetic order.
The resulting
antiferromagnetic state, with spin moments of $\approx$ 0.3 $\mu_B$/Os,
has the same energy as the paramagnetic state
to within the precision of our calculations. Furthermore, with this size of
moments, the band narrowing is insufficient to destroy the metallic state.
The Os sublattice consists of corner sharing tetrahedra, with the Os
atoms at the corners. As a result it is not possible to find a structure
in which all nearest neighbor Os bonds are antiferromagnetic,
and additionally,
the lattice is strongly geometrically frustrated for nearest neighbor spin
Hamiltonians. However, based on our band structure results, the proximity
of Cd$_2$Os$_2$O$_7$ to magnetism is itinerant in nature, and therefore
cannot be described in terms of simple nearest neighbor spin Hamiltonians.
In fact, the experimental low thermodynamic
data do not show any evidence of a large near ground state degeneracy.
However, it is still the case that the simple collinear antiferromagnetic
state we considered is favored relative to a ferromagnetic state, and
so one may speculate that more complicated arrangements that would better
fulfill a tendency for antiferromagnetic neighbors may be lower in energy.
In particular, even though small moment itinerant magnets are generally
collinear, as may be understood from arguments based on band kinetic
energy considerations, here for want of a better alternative, one might
postulate that a non-collinear ground state may occur. One
example of such a state would be an arrangement in which the Os moments
are directed either towards or away from the center of the tetrahedra.
Such a state could be at the $\Gamma$ point or could have the character
of a spin density wave, with modulation at a non-zero wavevector.

Each Os atom has two like spin and four opposite spin neighbors,
in the simple antiferromagnetic configuration we calculated.
On the pyrochlore lattice, this leaves a fully connected 3D
network of like spin nearest neighbor atoms. One may speculate that
a more complex arrangement that disrupts this fully connected network
(as the non-collinear arrangement mentioned above would do) would
narrow the bands enough to produce an insulating state.
It is unclear
whether such a state would be energetically favored here. However, the
fact that we find the simple collinear antiferromagnetic state to be at
least marginally stable in the GGA and favored with respect to ferromagnetism,
may suggest that it is at least possible. In an itinerant magnetic state,
like the one discussed above, the magnetic interactions in real space are
inherently long range, so depression of the ordering 
temperature, $T_N$ due to geometric frustration is not to be expected.
Still we emphasize that this scenario is not well
supported by the current results for three
reasons: (1) It is unclear that a purely magnetic instability exists
in the LSDA, which may be a better approximation than the GGA
for 5$d$ compounds. An alternate scenario is a transition involving
both charge and spin ordering, though there is not presently evidence
for a charge density wave in the experimental data; (2) Non-collinear
states, as mentioned,
are generally not found or expected in itinerant low moment magnets;
and finally (3) even if the system has magnetic instabilities it is
unclear that they could be
strong enough to produce a transition temperature
above 200K.

\section{Summary, Discussion and Conclusions}

Band structure calculations for Cd$_2$Os$_2$O$_7$ and Cd$_2$Re$_2$O$_7$
show a considerable
sensitivity of the electronic structure
near $E_F$ to spin orbit interactions.
Both materials show semi-metallic band structures with heavy bands
near $E_F$.
Cd$_2$Re$_2$O$_7$ has heavy hole bands near the
zone boundary and relatively light electron pockets around $\Gamma$,
while the electronic structure of
Cd$_2$Os$_2$O$_7$ is dominated by heavy bands for both the holes and
electrons. Interestingly, because of the higher $N(E_F)$ in Cd$_2$Os$_2$O$_7$,
the transport function, $Nv^2$ differs by only 7$\%$.
However, the specific heat enhancement in Cd$_2$Os$_2$O$_7$
is apparently quite small, while that in Cd$_2$Re$_2$O$_7$ is 2.4,
leaving little room for electron-phonon interactions or simple many electron
effects in the former. Furthermore, while Cd$_2$Os$_2$O$_7$ is
at least near antiferromagnetism in the GGA, neither compound shows
a clear proximity to magnetism in the LSDA, which may be a more reasonable
approximation for 5$d$ compounds. There are no clear nesting features
in the band structures that would suggest spin or charge density wave
Fermi surface instabilities.

It is tempting to speculate that the metal-insulator transition
in Cd$_2$Os$_2$O$_7$ and the 200K transition in Cd$_2$Re$_2$O$_7$
are related
to some common feature in their electronic structures.
In this regard, we note that although they have different electron counts,
both compounds have semimetallic band structures each
with nominally equal numbers
of holes and electrons dominated by very heavy bands.
This is suggestive of a excitonic instability of the
Fermi surface of the type proposed by Mott \cite{mott}
and reviewed by Halperin and Rice. \cite{halperin}
The theory in its simplest form involves pairing between
electrons on one sheet of Fermi surface and holes on another.
Conditions that favor such a state are (1) heavy band masses
(2) low carrier densities and (3) similar sizes and shapes of Fermi surfaces,
although the
latter condition can sometimes be
relaxed as discussed in Ref. \onlinecite{halperin}.
Such an instability is purely electronic, and
in contrast to {\it e.g.} a charge density wave need not be
coupled significantly to lattice degrees of freedom.
The excitonic state may be either singlet, or triplet.
Furthermore it can coexist with superconductivity 
\cite{ginzburg,mattis,kopaev}
or band ferromagnetism, \cite{volkov,ovchinnikov,zhitomirsky}
provided that there are excess carriers.
While such a transition is due to electron correlations, it is not
associated with an on-site Hubbard-like Coulomb repulsion, but rather
{\bf k} dependent correlations near $E_F$.
Importantly, the formation of an excitonic
state involves weak thermodynamic signatures
(it involves electrons only within a distance determined by the exciton
binding energy from $E_F$) and is continuous.

The general phenomenology of an excitonic metal insulator transition would
seem to fit experimental knowledge and the above band structure results
for Cd$_2$Os$_2$O$_7$, especially for triplet pairing. \cite{pairing-note}
The key parameters determining whether such a state can be formed are
the exciton binding energy and the
degree to which the size and shape of the electron and hole Fermi surfaces
match in {\bf k} as compared to the inverse size of an exciton. We cannot
assess these as we do have the effective dielectric screening.
However, we note that the very heavy bands should help to produce small
excitons and favor this possibility. The situation in Cd$_2$Re$_2$O$_7$
is similar. However, in this material the 200K transition is between
two metallic states. Within the excitonic scenario the 
metallic conduction below 200K and the superconductivity
depend on the existence
of non-paired carriers below the transition. These could be provided by
{\it e.g.} slight off-stoichiometry, which may perhaps be anticipated
from the unusual Re valence. This doping
could be sample dependent providing an
explanation of the very different $\partial\rho / \partial T$ and the
different $T_c$ of
the two reported samples discussed above.

We emphasize that the above discussion of excitonic states in
Cd$_2$Os$_2$O$_7$ and Cd$_2$Re$_2$O$_7$ is speculative. However,
if it is so, these materials will be interesting novel tests of
many body theories of excitonic phases, including superconductivity, 
possible
triplet pairing and the presence of strong spin-orbit interactions.

\acknowledgments

We are grateful for helpful discussions with D. Mandrus, W.E. Pickett
and S.G. Ovchinnikov.
Computations were performed using facilities of the DoD HPCMO ASC center.
Work at the Naval Research Laboratory is supported by the
Office of the Naval Research. J.O.S. is supported by CONICET Argentina.

\begin{figure}[tbp]
\centerline{\epsfig{file=dos-so.epsi,angle=270,width=0.95\linewidth}}
\setlength{\columnwidth}{3.2in} \nopagebreak
\caption{
LDA paramagnetic electronic density of states of Cd$_2$Os$_2$O$_7$
for the $t_{2g}$ manifold on a per unit cell (4 Os atoms)
basis. Spin-orbit is included.}
\label{ldados}
\end{figure}

\end{multicols}

\begin{references}
\bibitem{lapwnote}
Highly converged calculations were performed with two separate codes,
WIEN-97, (ref. \onlinecite{blahalapw})
and an independent general potential LAPW code, (ref.  \onlinecite{weilapw})
using independent choices of sphere radii,
zone integration and other parameters. No significant differences
were found in parallel calculations.
\bibitem{conv}
For the electronic structures, the LAPW sphere radii for which
projections of the density of states and other quantities were done
are 2.40 $a_0$, 2.05 $a_0$ and 1.55 $a_0$ for Cd, Os and O respectively.
\bibitem{hub-note}
It should be noted that a strong correlated system with an integer
filling would tend to be a Mott-Hubbard insulator both above and
below the magnetic ordering temperature.
\end{references}
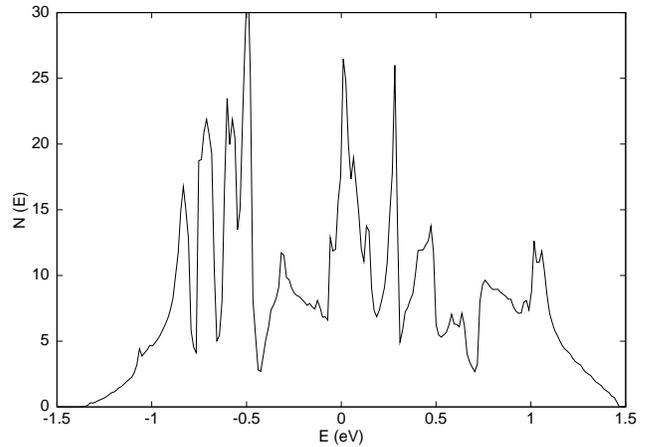
\end{document}